\documentclass[11pt,twocolumn]{IEEEtran}
\usepackage{authblk}
\usepackage{srcltx}
\usepackage[psamsfonts]{amsfonts}
\usepackage{amsmath}
\usepackage{amssymb}
\usepackage{graphics}
\usepackage{psfrag}
\usepackage{epsfig}
\usepackage{array}
\usepackage{algorithm}
\usepackage{algpseudocode}
\usepackage[psamsfonts]{amsfonts}
\usepackage{makeidx}  
\usepackage{bbm}
\usepackage{amsfonts}
\usepackage{hhline}
\usepackage{eufrak}
\usepackage{yfonts}
\usepackage{pifont}
\usepackage{footnote}
\usepackage{pdfpages}
\usepackage{graphicx}
\usepackage{balance}
\usepackage{color}
\usepackage[square, comma, sort&compress, numbers]{natbib}
\usepackage{dsfont}
\usepackage{caption}
\usepackage{subcaption}
\usepackage{sidecap}
\usepackage[splitrule,hang,marginal]{footmisc}
\usepackage{enumitem}

\newcommand{\lk}{ \left\{ }
\newcommand{\rk}{ \right\} }

\newcommand{\db}{{\bf d}}

\newcommand{\xb}{{\bf x}}
\newcommand{\rb}{{\bf r}}
\newcommand{\yb}{{\bf y}}

\newcommand{\Db}{{\bf D}}

\newcommand{\Xb}{{\bf X}}
\newcommand{\Yb}{{\bf Y}}

\newcommand{\no}[1]{\|#1\|_1} 
\newcommand{\nf}[1]{\|#1\|_F} 
\newcommand{\nt}[1]{\|#1\|_2}

\newcommand{\ub}{{\mathbf u}}

\begin{document}
 \pagenumbering{gobble}
\title{Functional Brain Networks Discovery Using Dictionary Learning with Correlated Sparsity }
\author{Mohsen Joneidi\\
Department of Electrical Engineering and Computer Science \\University of Central Florida\\
mohsen.joneidi@ucf.edu\\
$\;$\\
COURSE PROJECT REPORT}

\vspace{-10mm}
\maketitle

\vspace{-4mm}
\textbf{Summary}:
Analysis of data from functional magnetic resonance imaging (fMRI) results in constructing functional brain networks. Principal component analysis (PCA) and independent component analysis (ICA) are widely used to generate functional brain networks. Moreover, dictionary learning and sparse representation provide some latent patterns that rules brain activities and they can be interpreted as brain networks. However, these methods lack modeling dependencies of the discovered networks. In this study an alternative to these conventional methods is presented in which dependencies of the networks are considered via correlated sparsity patterns. We formulate this challenge as a new dictionary learning problem and propose two approaches to solve the problem effectively.

\section{Motivation}
Identifying brain networks and their interactions require the analysis of the recorded signals over time. The correlation of functional brain networks (FBNs) extracted from fMRI is a powerful tool for diagnostic purposes. fMRI is based on blood oxygenation level-dependent (BOLD) contrast, that is captured using local fluctuations in the flow of oxygenated blood. To perform different brain tasks, specific brain functional networks might be activated. They will be engaged collaboratively to execute a specific task. These networks are related to low-level brain functions, called segregated specialized small brain regions \cite{perlbarg2008contribution}. These regions collaborate to perform tasks; however, a few regions will be activated  to execute a certain task. Activation of a small number of brain functions for each task implies a kind of sparsity in terms of the fundamental functional bases. The recent promising results of dictionary learning (DL) based fMRI signal analysis confirm that the assumed sparsity is consistent with the pattern of brain activities. DL aims to decompose the observed signals in terms of some fundamental bases and their corresponding sparse coefficients. This method beats the traditional methods including principal component analysis (PCA) and independent component analysis (ICA) for extraction of the fundamental bases of activity patterns \cite{zhao2015supervised,lv2015sparse,ramezani2015joint}. The traditional methods assume \emph{orthogonality} or \emph{independency} for the fundamental bases which is an unnatural constraint for the  bases \cite{mckeown1998independent, daubechies2009independent, lee2011data}.              

DL is a promising alternative of ICA for fMRI signals decomposition. Recent studies have shown that DL outperforms ICA in this application due to more relaxed assumptions on the bases \cite{eavani2012sparse}. Moreover, it assumes a more flexible model for data. Additionally, the underlying model of DL is consistent with sparse activation of FBNs. It means only few FBNs are active in each instant. In addition to temporal sparsity of FBNs, the spatial activity patterns also are sparse \cite{papo2014reconstructing}. However, there exists a correlation between spatial activation patterns in the spatial domain. It  means the non-zero entries are not spread element-wise and there is a piece-wise sparsity which cause smoothness in FBNs. These characteristics motivate us to impose the dictionary learning model for data extracted from brain. Atoms of the dictionary span a union of reliable subspaces which is fitted to the training data. Wavelet decomposition, compressive sensing and DL are some products of union of subspaces model \cite{AharEB06,eldar2009robust}.

fMRI data consists of a collection of time series for each brain's voxel. These time series can be represented in terms of a collection of principal time series. However, the number of principal time series is limited. The fluctuation of each voxel generates a certain pattern in time and this time-series can be represented in terms of the brain's principal time series. Different brain regions collaborate to perform a certain function, while execution of a certain brain function only implies a few principal time series to be involved. Therefore, there exists an inherent  sparsity for representation of the signals of brain activity. 

The general model of DL assumes activation of the bases are independent of each other, i.e., activation of a principle basis does not effect on activation of any other bases. We call this type of sparsity as \emph{element-wise} sparsity. In addition to element-wise sparsity, some \emph{structural sparsity} also exists in the pattern of brain activities. Lack of regulation constraints such as orthogonality and coherency for the basic time series causes high-sensitive set of atoms.  a merely sparse constraint results in inconsistent activation maps. To address this problem, we exploit unsupervised group sparsity constraints for dictionary learning. Our goal is to engage the correlation of the principal time-series on their corresponding sparse coefficients.
This paper has two main contributions. First, proposing a new dictionary learning problem for single subject fMRI resting state signals decomposition, and second, presenting two computational algorithms for solving the proposed problem.

The rest of the paper is organized as follows. Section II reviews the related work in the literature. Section III presents some basic methods for estimation in terms of correlated bases. Inspired by Section III, a new dictionary learning problem is introduced in Section IV. Experimental results are presented in Section V and Section VI concludes the paper.  

\section{Related Works}

\begin{figure}
\centering
\includegraphics[width=3.5 in]{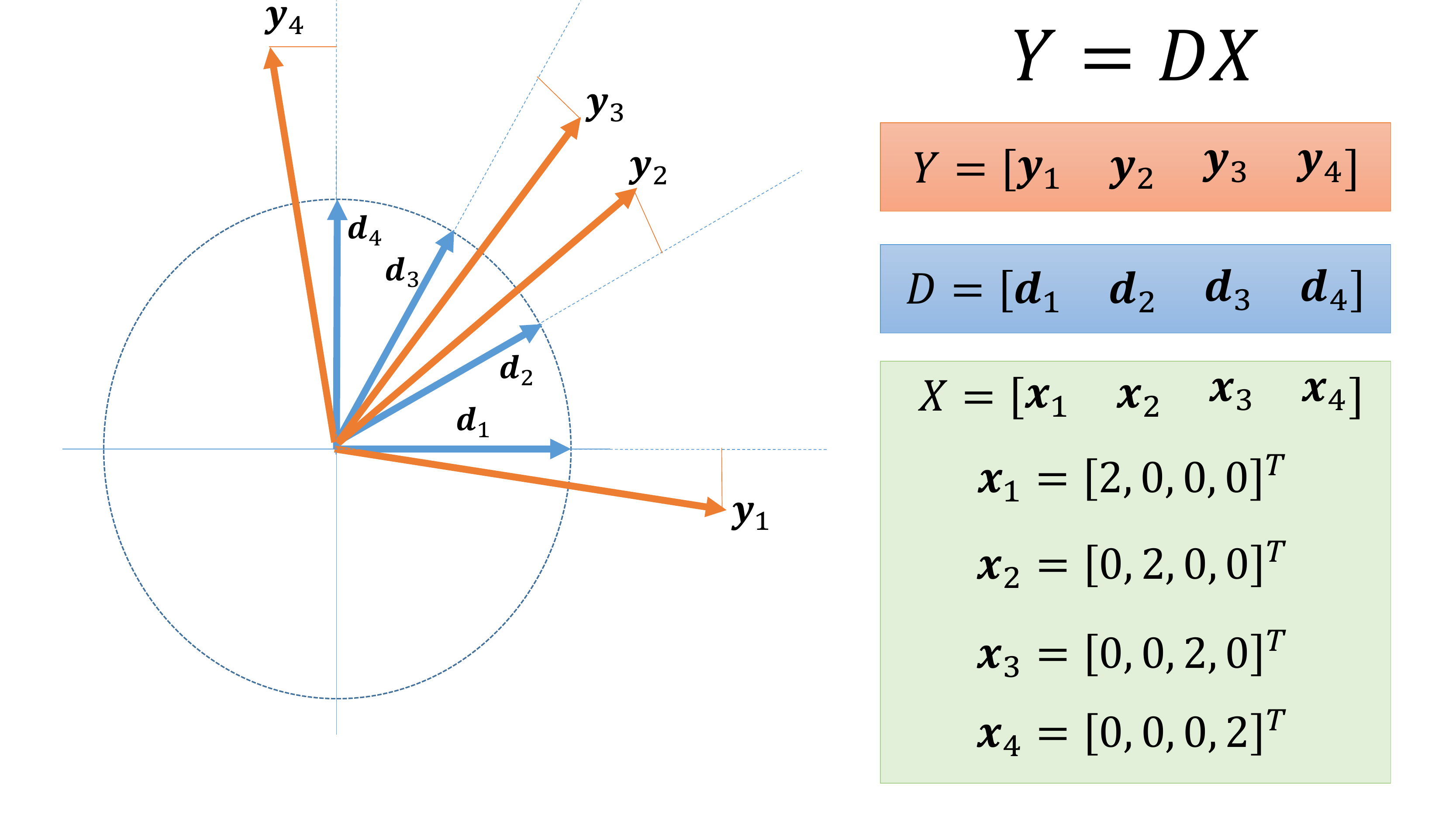}
\vspace{-3mm}
\caption{A simple example illustrates the inconsistency of element-wise sparsity.}
\label{incons_ex}
\end{figure}

Multivariate statistical algorithms consider brain voxels' activity as a collaborative network and analyze the voxels' data jointly. These methods include PCA, factor analysis and ICA \cite{ma2007detecting}. ICA is a data-driven method that can decompose the observed multivariate data into some maximally independent sources. This method does not need prior information neither on the sources nor on the characteristics of the mixing. ICA has been widely used to reveal brain networks for both task-based and resting state fMRI signals \cite{lee2013resting}. Although ICA exhibits a relatively fit model for fMRI signals, it cannot impose additional prior information such as sparsity. There exist many efforts to consider sparsity for fMRI signals analysis \cite{yamashita2008sparse, lee2011data}. DL is a sparsity-based method that has received much attention recently. The basic dictionary model is given by,
\begin{equation}
\label{eq:model}
\Yb = \Db \Xb +\mathbf{N}, \quad \|\mathbf{x}_i \|_{0}\le T,
\end{equation}
in which $\Yb \in \mathbb{R}^{N\times L}$ is the observed signals, $\Db \in \mathbb{R}^{N\times K}$ is the dictionary and $\Xb \in \mathbb{R}^{K\times L}$ is the coefficients matrix, $N$ is the dimension of the observed data, $L$ is the number of observed data and, $K$ is the number of bases in the dictionary. $\mathbf{x}_i$ is the $i^{\text{th}}$ column of $\Xb$ and $\|.\|_0$ denotes $\ell_0$ norm which counts the number of nonzero elements of a vector. $T$ is the maximum number of nonzero entries in each column. Each column of $\Yb$ contains an observed signal, and the corresponding column in $\Xb$ is its sparse representation. The basic DL can be cast as an optimization problem as follows,
\begin{equation}
\label{eq:dl}
\min_{\Xb,\Db}~\frac{1}{2}\nf{\Yb-\Db\Xb}^2\quad \text{s.t.} \;\|\xb\|_0\le T.
\end{equation}
A well-known solution for this problem is the K-SVD algorithm which is an alternative optimization method. I.e., it initializes $\Db$ and optimizes w.r.t $\Xb$ and $\Db$ alternatively \cite{AharEB06}. Optimization w.r.t $\Xb$ is called \emph{sparse coding} and optimization w.r.t $\Db$ is called dictionary updating. Sparse representation of $\Yb$ is leaned in $\Xb$, however this representation can be learned such that provides a discriminative representation for $\Yb$ when input data are labeled \cite{golmohammady2014k}. K-SVD is used for resting state and task-based fMRI signal analysis. The only constraint on the dictionary is the normalization of its columns. K-SVD does not consider any additional constraint on the dictionary. However, in presence of highly correlated bases the corresponding coefficients are not consistent. In other words distances in the original data space are not projected to the sparse representation space consistently. Fig. \ref{incons_ex} illustrates this effect. Suppose we are given $4$ bases, $\textbf{d}_1,\cdots, \textbf{d}_4$, distributed on the unit ball. Some data, $\textbf{y}_1,\cdots, \textbf{y}_4$ are approximated by the underlying bases and their corresponding sparse representations are denoted as $\textbf{x}_1,\cdots, \textbf{x}_4$. Only one basis is utilized to represent each data, i.e., the sparsity is 1. Obviously, the distance between each pair of data is so much variant while the distance between each pair of sparse vectors is $2\sqrt[]{2}$. It means a small change in data space may cause a big abrupt change in the sparse domain. Correlation between atoms of a dictionary should be considered in order to estimate a more accurate sparse representation \cite{joneidi2015matrix}.

In order to alleviate the inconsistency of sparse representation, imposing an incoherency constraint for the dictionary bases could be a remedy that suggested in \cite{abolghasemi2015fast} for analyzing fMRI signals. This idea aims to learn incoherent dictionary bases by considering a new regularizer term.
\begin{equation}
\label{eq:low_coh}
\min_{\Xb,\Db}~\frac{1}{2}\nf{\Yb-\Db\Xb}^2+\lambda \|\Xb\|_1 +\gamma \|\Db^T \Db -I\|_F^2.
\end{equation}
where, $\lambda$ and $\gamma$ encourage sparsity and incoherency, respectively. The last term adds a penalty for the off-diagonal elements of the correlation matrix $\Db^T\Db$. However, in this paper we do not assume any additional constraint on the dictionary but the sparse coding is modified in a way to compensate the destructive effect of highly correlated bases of the dictionary on the sparse coefficients. In the next section the undesired effect of correlated bases is discussed and some existing solutions are explained. 
\vspace{-3mm}
\section{Dealing with Correlated Bases}

In some applications, atoms of the dictionary may be highly correlated. In this setting, we expect the coefficients corresponding to correlated atoms to be correlated with each other. For example, if two atoms are highly correlated, then they would be either both zero or both non-zero. Thus, the coefficients tend to appear in groups, i.e., the coefficients corresponding to a subset of correlated atoms are all together equal to zero, or they are all non-zero. This is known as the grouping effect which is of a high importance especially in linear regression \cite{ogutu2012genomic}. 

Traditional $\ell_1$-regularization, as used in LASSO \cite{tibshirani1996regression}, though lead to sparse coefficients, it fails to maintain the grouping effect in highly correlated bases. That is, from the coefficients corresponding to the same group of correlated atoms, only one of them may be non-zero due to the crisp sparsity behaviour of the $\ell_1$-norm. To remedy this problem, group sparse regularizing terms have been proposed \cite{meier2008group, efron2004least, zou2005regularization}. The group structure is considered in both predefined model-based and unknown data-driven manners. One such interesting data-driven solution has been proposed by Zhou et al, by the elastic net regularization \cite{zou2005regularization}. The elastic net term comprises of a sparsity term, the $\ell_1$-norm, plus a grouping and stabilizing term, the $\ell_2$-norm. That is, elastic net merges the benefits of the LASSO and the ridge regression \cite{ogutu2012genomic}. The final sparse coding (regression) problem is then,
$$ \min_{\xb}~\lk\frac{1}{2}\nt{\yb-\Db\xb}^2+\lambda_1\no{\xb}+\lambda_2\nt{\xb}^2\rk.$$

The grouping effect is essential in the brain functional analysis, where there may exist strong correlations among various activity areas in the brain. 

\section{Dictionary Learning with Correlated sparsity Constraints}
\begin{algorithm}[b]
\caption{OMP sparse coding}
\label{alg:omp}
\begin{algorithmic}[1]
\State \textbf{Require:} $\yb$, $\Db$, $\epsilon$.
\State \textbf{Initialization:} $\rb^0=\yb, \mathbb{S}=\{\}, \xb=\textbf{0}$ and $i=0$.
\While{$\mbox{Error}>\epsilon$}
\State $\mathbb{S}=\mathbb{S}\bigcup \underset{j}{\text{\small{argmax}}} <\rb^i,\db_j>$
\State $\xb_\mathbb{S}= \underset{\xb}{\text{argmin}} \|{\yb - \Db({:,\mathbb{S}})\xb}\|_2^2$
\State $i\leftarrow i+1$
\State $\rb^i=\yb-\Db\xb$
\State $\mbox{Error}=\|{\rb^i}\|_2$
\EndWhile
\State \textbf{Output:} $\xb$
\end{algorithmic}
\end{algorithm}

\begin{algorithm}[b]
\caption{Updating Dictionary's atoms \cite{AharEB06}}
\label{alg:dic_up}
\begin{algorithmic}[1]
\State \textbf{Require:} $\Yb$, $\Xb$.
\State \textbf{for $k=1,\cdots,K$} 
\State \quad Collect all the data that use $\db_k$ in $\mathbb{S}_k$ set
\State \quad Assume $\db_k=\textbf{0}$ and compute $\boldsymbol{E}_k=\Yb-\Db\Xb$
\State \quad Reduce $\boldsymbol{E}_k$ by $\mathbb{S}_k$ columns$\rightarrow \boldsymbol{E}_k^R$ 
\State \quad Update $\db_k$ by the first left Eigenvector of $\boldsymbol{E}_k^R$ 
\State \textbf{end}
\State \textbf{Output:} $\Db$
\end{algorithmic}
\end{algorithm}

Plain DL algorithms like KSVD are not able to learn incoherent  bases. Moreover, their coefficients are not consistent in presence of correlated bases. This is more crucial in the brain network analysis because the functional brain networks are not independent \cite{mckeown1998independent, daubechies2009independent, lee2011data}. Herein, we propose an elastic-net-based \cite{zou2005regularization} dictionary learning formulation to solve this problem. Herein, our proposed problem is as follows,
\begin{equation}
\label{eq:elnet}
\underset{\Xb,\Db}{\text{argmin}}~\frac{1}{2}\nf{\Yb-\Db\Xb}^2+\lambda\mbox{EN}(\Xb),
\end{equation}
where,
\[ \mbox{EN}(\Xb)\triangleq \no{\Xb}+\frac{\gamma}{2}\nf{\Xb}^2,\]
is the matrix form of the elastic-net regularization. We use the proximal-spliting algorithm \cite{PariB14} to solve Problem \eqref{eq:elnet}. Proximal-spliting method targets the following optimization problem:
\[ \underset{\xb}{\text{argmin}}~\lk f(\xb)=g(\xb)+h(\xb) \rk, \]
where $g(.)$ and $h(.)$ are convex functions with $g(.)$ being differentiable in addition. The idea is then to perform the following iterations to update $\xb$
\begin{equation}
\xb_{k+1}=\mbox{Prox}_h(\xb_k-\mu\nabla{g}(\xb_k)).
\end{equation}
where $\mbox{Prox}_h(.)$ is the so-called proximal operator of $h(.)$ defined as
\[ \mbox{Prox}_h(\xb)\triangleq \underset{\ub}{\text{argmin}}~\lk\frac{1}{2}\nt{\xb-\ub}^2+h(\ub)\rk \]
Also, the step size $\mu\in(0,\frac{1}{L}]$, in which $L$ is the Lipschitz constant of $g(.)$.

In problem \eqref{eq:elnet}, we have
\[ g(\Xb)=\frac{1}{2}\nf{\Yb-\Db\Xb}^2,~h(\Xb)= \lambda\mbox{EN}(\Xb)\]
It can be shown that $L=\| \Db^T\Db \|$, where $\| . \|$ denotes the spectral norm. The proximal operator of the elastic net term has also a simple close-form formula
\begin{equation}
\mbox{Prox}_{\mbox{EN}}(\Xb)=\frac{1}{1+\lambda\gamma}\mbox{Soft}(\Xb, \lambda)
\end{equation}
where $\mbox{Soft}(.,.)$ is the well-know element-wise soft-thresholding function defined as
\[ \mbox{Soft}(\Xb, \lambda)\triangleq\mbox{sign}(\Xb)\odot \max(|\Xb|-\lambda,0). \]
In which, $\odot$ indicates element-wise multiplication. Our proposed scheme contains two main subroutines as same as conventional DL algorithms, sparse coding and dictionary updating. The steps of these subroutines are explained in Alg.~\ref{alg:omp} and Alg.~\ref{alg:dic_up}. Moreover, the overall iterative sparse coding algorithm is summarized in Alg.~\ref{alg:elnet}. In this algorithm, we use the relative change between consecutive solutions of the iterative algorithm as a stopping criterion. This sparse coding is then used as the sparse approximation stage of our proposed dictionary learning algorithm. For the dictionary update stage, any algorithm like KSVD atom-by-atom dictionary update can be used. Moreover, for initializing the elastic-net sparse coding we leverage the iterative nature of the dictionary learning problem, and use the coefficient matrix $\Xb$ of the previous DL iteration as a warm start. This greatly reduces the computational burden of the sparse coding stage.

\begin{algorithm}[t]
\caption{Elastic-Net Regularized DL}
\label{alg:elnet}
\begin{algorithmic}[1]
\State \textbf{Require:} $\Yb$, $\lambda$, $\gamma$, $\epsilon$
\State \textbf{Initialization:} $\Db=\Db_0\in\mathbb{R}^{N\times K}$, $\xi=\infty$, $\mu=1/\| \Db^T\Db \|$
\While{stopping criterion is not met}
\State \textbf{Sparse approximation} (Alg. \ref{alg:omp})
\While{$\xi>\epsilon$}
\State $\Xb_o=\Xb$
\State $\Xb=\Xb-\mu\nabla g{\Xb}$ \Comment{Gradient step}
\State $\Xb=\frac{1}{1+\lambda\gamma}\mbox{Soft}(\Xb, \lambda)$ \Comment{Proximal mapping}
\State $\xi=\nf{\Xb-\Xb_o}/\nf{{\Xb_o}}$
\EndWhile
\State \textbf{Dictionary update} (Alg. \ref{alg:dic_up})
\EndWhile
\State \textbf{Output:} $\Xb$ and $\Db$
\end{algorithmic}
\end{algorithm}
\begin{algorithm}[b]
\caption{Modified Grouped-wise K-SVD}
\label{alg:mod_ksvd}
\begin{algorithmic}[1]
\State \textbf{Require:} $\Yb$, $\lambda$, $\epsilon$ .
\State \textbf{Initialization:} $\Db=\Db_0\in \mathbb{R}^{N\times K}$
\While{$\mbox{stopping criterion is not met}$}
\State $\textbf{G}=\mbox{HT}(\Db^T \Db,\lambda)$ 
\State \textbf{Sparse Approximation} (Alg. \ref{alg:omp})
\State $\textbf{X} \;	\leftarrow \; \textbf{G}\Xb $ 
\State   $\textbf{X}^{*}	\leftarrow \; \Xb\Sigma$  using Eq. (\ref{eq:nrml})
\State \textbf{Dictionary Update} (Alg. \ref{alg:dic_up}) 
\EndWhile
\State \textbf{Output:} $\Xb^{*}$
\end{algorithmic}
\end{algorithm}
To prove the efficiency of the proposed grouped variables approach in DL problem, we have simply modified the K-SVD DL algorithm. To this aim, in the sparse coding stage the correlations of the dictionary bases are taken into account to construct a set of grouped-wise coefficients. $\textbf{G}$ is defined as hard-threshold (HT) of the bases correlations. HT function is illustrated in Figure 1. Note that if the bases are low-correlated, $\textbf{G}$ will be close to the identity matrix and the algorithm works like the basic K-SVD. However, in the case of highly correlated bases, multiplication of $\textbf{G}$ by the coefficients results in group sparsity of the highly correlated coefficients. At the end the new coefficients should be normalized to minimize the reconstruction error. The scale can be calculated easily,
\begin{equation}
\Sigma_{ii}=\underset{\sigma}{\text{min}} \|\yb_i-\sigma \Db \xb^{*}_i\|_2^2= \frac{\yb_i^{T}\Db\xb^{*}_i}{\xb_i^{*T}\Db^T\Db\xb^{*}_i}.
\label{eq:nrml}
\end{equation}

In which, $\boldsymbol{\Sigma}$ is a diagonal normalization matrix.
Algorithm 2 shows the steps of the modified algorithm. In this algorithm only the coding is modified and the dictionary update stage is remained like the original K-SVD algorithm.

\begin{figure}[t]
\centering
\vspace{-2mm}
\includegraphics[width=2 in,angle=0]{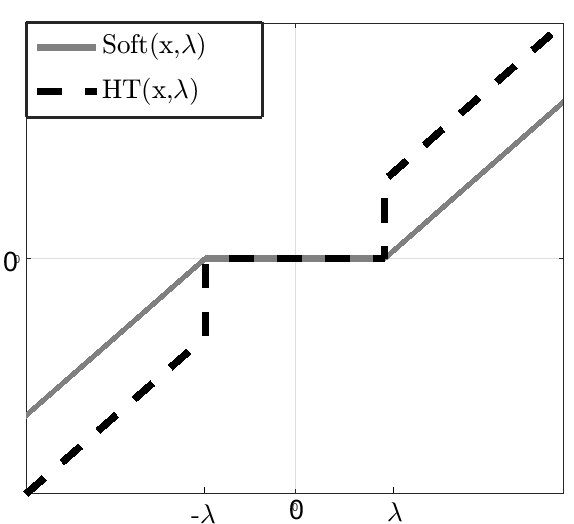}
\caption{\small{Soft threshold versus hard threshold.}}\label{soft_ht}
\end{figure}

\section{Experimental Results}
The proposed algorithms are evaluated in two simulation scenarios, synthetic and real fMRI data.
\begin{figure}[b]
\begin{subfigure}{0.5\textwidth}
\centering
\includegraphics[width=2.5 in]{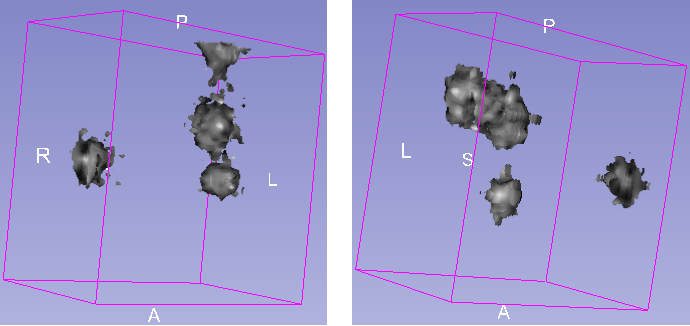}
\caption{Two synthetic principal bases of FBNs.}
\end{subfigure}
\begin{subfigure}{0.5\textwidth}
\centering
\includegraphics[width=2.5 in, height=1.2 in]{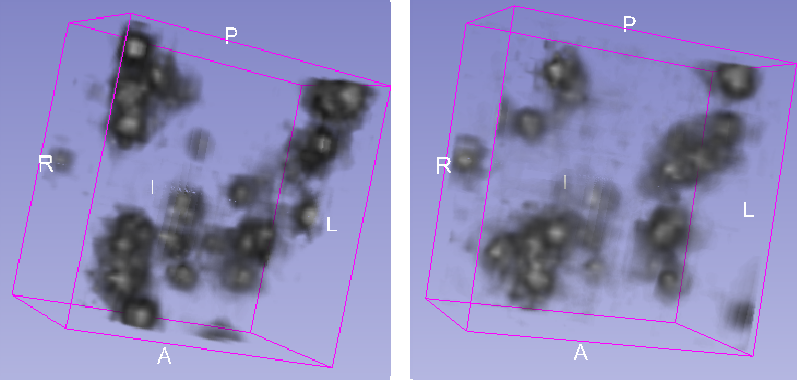}
\caption{Synthetic fMRI data in two time courses obtained by modulation of 10 FBNs by their corresponding time series with additive Gaussian noise.}
\vspace{-3mm}
\label{synthetic}
\end{subfigure}
\caption{\small{Illustration of synthetic fMRI data generation.}}
\end{figure}
\subsection{Synthesized fMRI Data}
This experiment examines the performance of proposed methods in separating the sources of some artificially generated fMRI data. Some 3D images are considered as functional brain networks and they are modulated by some principle time series and an additive Gaussian noise is added to generate the final synthetic data. Figure 2 shows some underlying brain networks and their activation time series as well as some generated 3D fMRI image.

To evaluate the obtained dictionary using different methods, the following dictionary distance is used \cite{AharEB06}.
\begin{equation}
{d}_d (\Db_{0}, \hat{\Db})=\sum_{k=1}^K \underset{j}{\text{min}} (1- \hat{\db}_k^T \hat{\db}_j^0)
\end{equation}

\begin{figure}[t]
\centering
\vspace{-2mm}
\includegraphics[width=3 in,angle=0]{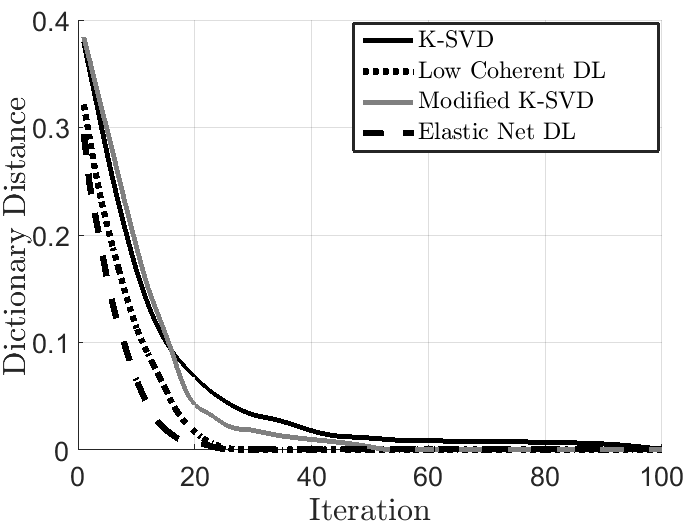}
\caption{\small{Performance of the different DL algorithms over iterations.}}\label{simul1}
\end{figure}

\begin{figure}[t]
\centering
\vspace{-2mm}
\includegraphics[width=3 in,angle=0]{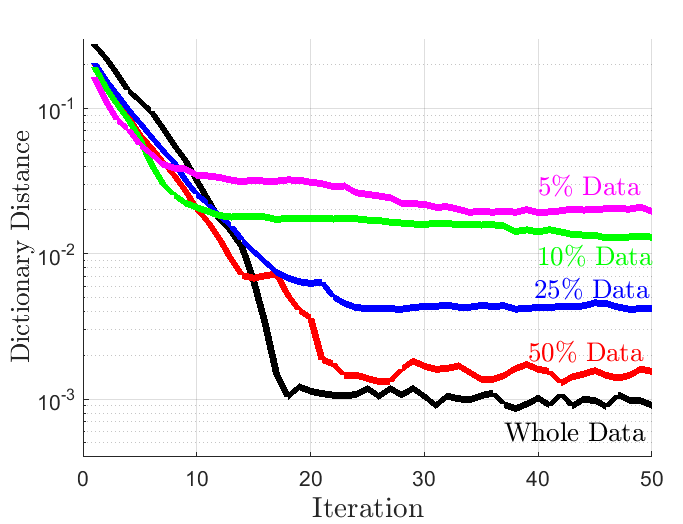}
\caption{\small{Performance of the proposed algorithm while it observes a portion of the whole data.}}\label{simul1}
\end{figure}

\subsection{Real fMRI}
A single subject analysis is performed to compare the traditional dictionary learning with the modified one. Resting-state fMRI data are downloaded from a free access online dataset\footnote{http://www.myconnectome.org}. SPM 12 Matlab toolbox is used to perform the needed pre-processing such as normalization and registration.  

\begin{figure}[b]
\centering
\vspace{-2mm}
\includegraphics[width=3 in,angle=0]{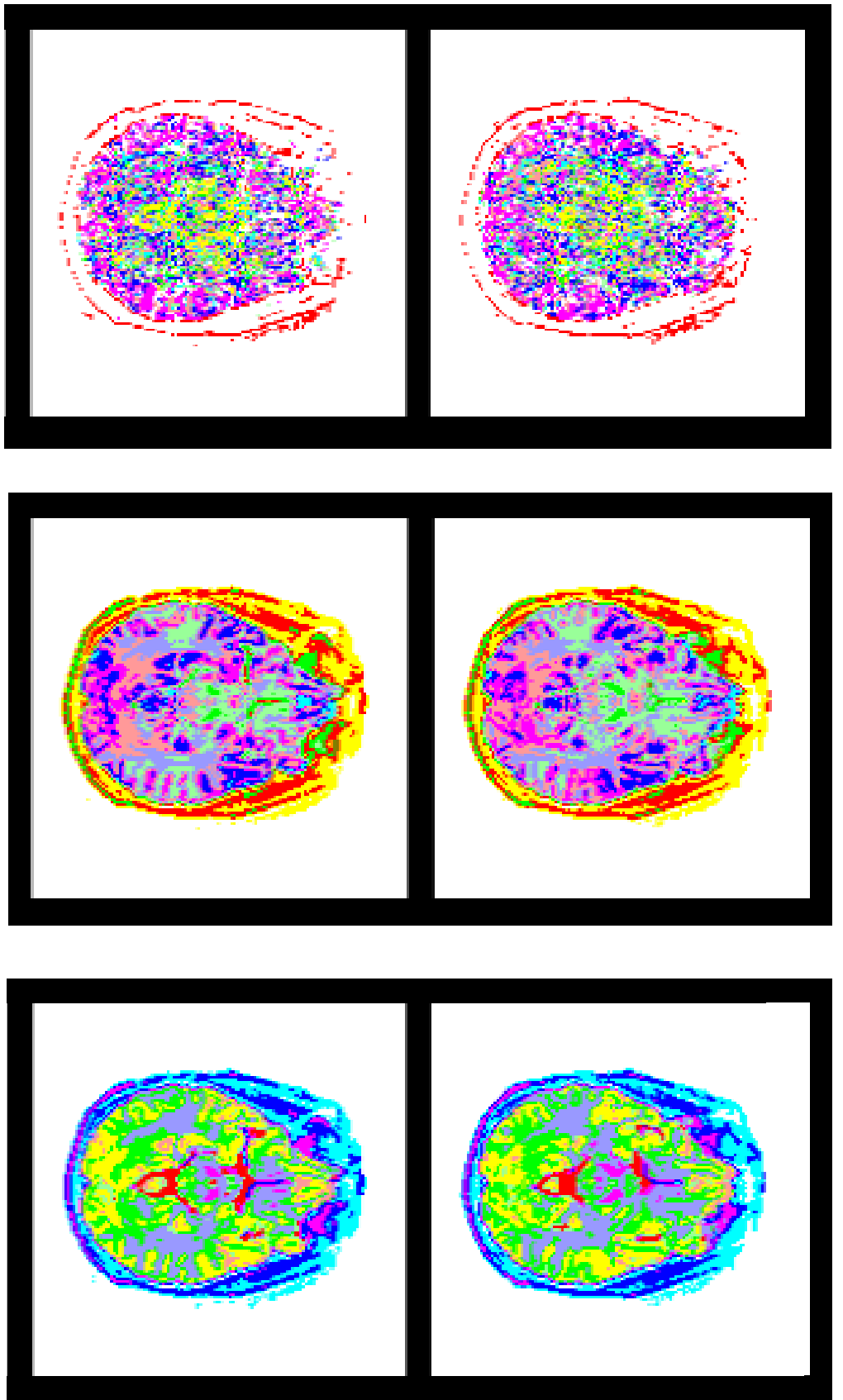}
\caption{\small{The segmentation results in two slices of brain using pure sparsity constraint (the upper image) versus proposed sparsity. The proposed sparsity is solved using two proposed algorithms. The middle one is the Modified-KSVD and the bottom image is resulted by the EN-KSVD. }}\label{fmri_accurate}
\end{figure}

\begin{figure*}[t]
\centering
\vspace{-6mm}
\begin{subfigure}{0.95\textwidth}
\centering
\includegraphics[width=3 in]{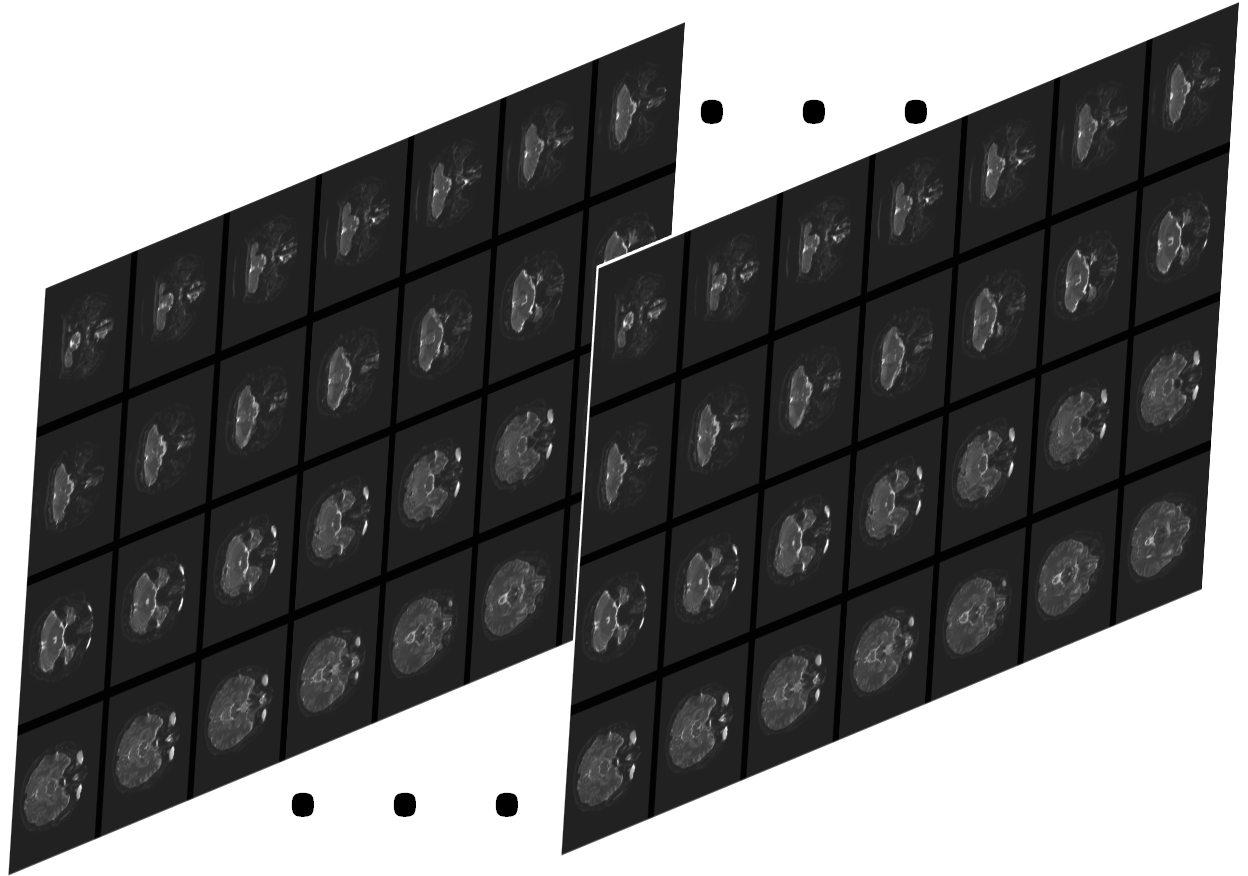}
\vspace{-1mm}
\caption{Resting state fMRI data in which 50 time slots are observed.}
\end{subfigure}
\begin{subfigure}{0.95\textwidth}
\centering
\includegraphics[width=5.6 in]{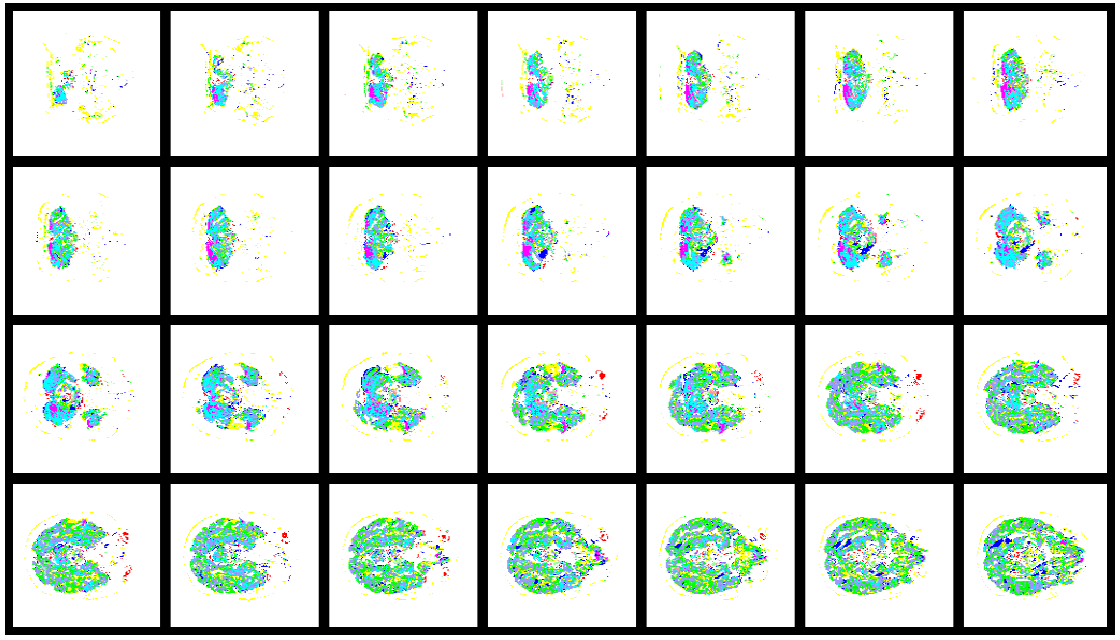}
\vspace{-1mm}
\caption{\small{Segmentation of region activities using K-SVD coefficients.}}
\end{subfigure}
\begin{subfigure}{0.95\textwidth}
\centering
\includegraphics[width=5.6 in]{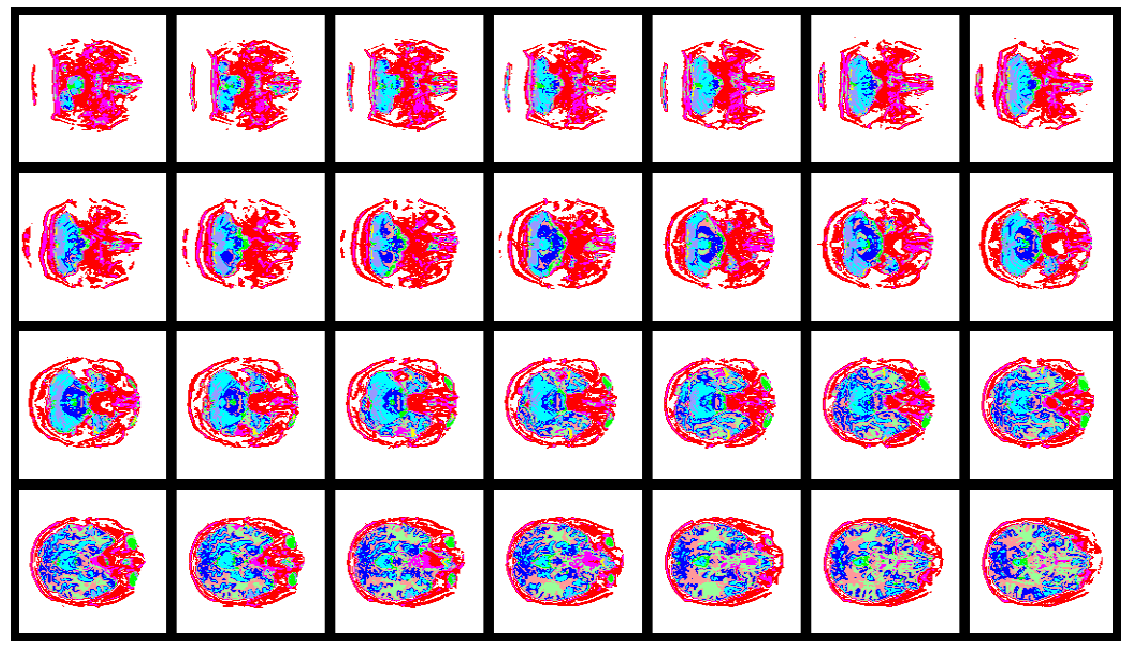}
\vspace{-1mm}
\caption{\small{Segmentation of region activities using the proposed DL coefficients.}}
\end{subfigure}
\vspace{-1mm}
\caption{Applying brain segmentation on coefficients extracted by the K-SVD and the proposed dictionary learning.}
\label{fMRI_cluster}
\end{figure*}

The spatial resolution for fMRI data is 160x160x36 pixels where we access to 50 time courses. Corresponding to each point there is a time series. All of these time series are collected as columns of $\boldsymbol{Y}$ matrix, then it is decomposed to some bases and coefficients. The coefficient of each point is exploited to perform segmentation using clustering. A simple clustering-based segmentation is performed by K-means with $\ell_1$ criterion. As it can be seen in Fig. \ref{fMRI_cluster}, the modified coefficients are able to segment the volume of the brain to result in  coherent regions which is consistent with functional brain networks. Fig. \ref{fmri_accurate} shows the segmentation of the $27^{\text{th}}$ and the $28^{\text{th}}$ slice of brain more accurately. The upper image shows segmentation using the pure sparsity features, the middle one indicates the obtained results using the proposed Modified-KSVD algorithm and the bottom image is resulted by the Elastic-net dictionary learning.     

\section{Conclusion}
Principal activity patterns of the brain are detected. Sparsity of the activation maps utilized in a framework based on dictionary learning. The correlated sparsity pattern of the underlying data showed advantage over pure sparsity pattern due to taking into account the dependencies of functional brain networks.

\footnotesize{
\balance

}
\end{document}